# Bridging the Research-Practice Gap in Requirements Engineering through Effective Teaching and Peer Learning

Andrew M. Connor, Jim Buchan & Krassie Petrova
*School of Computing & Mathematical Sciences, Auckland University of Technology*

**Abstract**

*In this paper, we introduce the concept of the research-practice gap as it is perceived in the field of software requirements engineering. An analysis of this gap has shown that two key causes for the research-practice gap are lack of effective communication and the relatively light coverage of requirements engineering material in University programmes. We discuss the design and delivery of a Masters course in Software Requirements Engineering (SRE) that is designed to overcome some of the issues that have caused the research-practice gap. By encouraging students to share their experiences in a peer learning environment, we aim to improve shared understanding between students (many of whom are current industry practitioners) and researchers (including academic staff members) to improve the potential for effective collaborations, whilst simultaneously developing the requirements engineering skill sets of the enrolled students. Feedback from students in the course is discussed and directions for the future development of the curriculum and learning strategies are given.*

***Keywords:*** Requirements engineering, research-practice gap

## 1. Introduction

The concept of a research-practice gap is not new, and it has been identified in many domains such as nursing [1], management [2] and marketing [3] to name but a few. Fitzgerald [4] has identified that in applied disciplines, such as information systems, the gap between theory/research and practice is a worrying one. In addition, Lang [5] argues that "because research findings often do not have direct or immediate relevance to IS professionals in industry, the question arises as to how those findings should be disseminated to them in a suitable form at such time as they do become relevant".

In this paper, we adopt an alternative view based on a previous analysis of the research-practice gap in requirements engineering [6]. Rather than address the relevance issues of research, we propose that the first step in bridging the research-practice gap is to gain a shared understanding of the issues that relate not just to different perspectives of requirements engineering as a domain, but also in terms of what practitioners and researchers can potentially achieve from collaborative working. To that end, we detail the design and delivery of a Masters level course in Software Requirements Engineering (SRE) that is specifically intended to address the needs of researchers and practitioners to understand each other in the field of Requirements Engineering. Such understanding is not only achieved by the students enrolled in the course. By observing the development of this shared understanding at the student level, research active academic staff can reflect on their own interactions with industry research partners as a means of developing more effective research partnerships. This course is currently delivered as part of the Master of Computer & Information Sciences (MCIS) degree at Auckland University of Technology. Before discussing the design of the SRE course and how it achieves the goal of bridging the research-practice gap (section 4) the philosophy, aims and structure of the MCIS degree are described in the next section. This provides a context for the SRE course and a profile of the students taking it. This is followed by a summary of the current understanding of the research-practice gap, specific to the domain of Requirements Engineering (RE), which have influenced and informed the design of the SRE course. In particular the main barriers to the bridging of this gap are highlighted. In section 5 the outcomes of the first two iterations of the SRE course are evaluated and discussed, and some evidence from student feedback provided. Some future directions for further work are suggested in the concluding remarks in section 6.

## 2. The MCIS structure

Postgraduate degrees offered at Auckland University of Technology (AUT) have traditionally had a strong emphasis on professional education and having only gained University status in 2000 there is a changing emphasis towards research development. As a result, it is important to attempt to balance research and professional practice. The Master of Computer & Information Sciences (MCIS) degree not only addresses the educational needs of Information Technology (IT) professionals, but also provides an infrastructure for growing research capability. The underpinning educational philosophy places a high value on providing a solid theoretical and research-based foundation from which best professional IT practice can be derived. The MCIS degree evolved from its predecessor the Master of Information Technology. Traditionally, the student body would have included IT professionals wishing to extend and update their technical and managerial skills or capabilities. These students are generally motivated to complete the degree and advance their career. Another segment of the student population is those students who have recently completed their undergraduate studies and are using the Master's degree as a vehicle to change career direction or gain a local (New Zealand) IT qualification. A small, but growing, segment includes those students wishing to gain entry to further (doctoral) studies and pursue a research based career. The structure of the MCIS programme is sufficiently flexible to accommodate all of these students. Figure 1 illustrates the structure of the degree, which includes two core courses, a range of electives and either a dissertation or thesis.

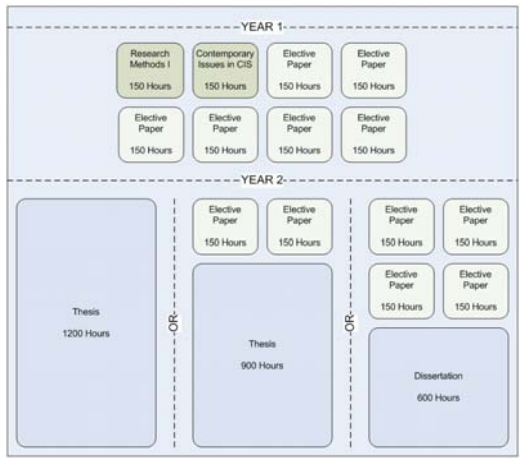

Figure 1. MCIS structure

The current list of elective courses on offer includes Net-centric Computing, Usage Centered Design, Ubiquitous Computing, Integrating IT & the Enterprise, ICT Issues in the SME Sector, Bioinformatics, Data Mining & Knowledge Engineering, Information Security, Computer Graphics and Animation, Health Informatics, Research Methods 2, Software Requirements Engineering, eSystems Design and Development, Data Warehousing, Artifical Intelligence as well as the provision for conducting either set Readings course or proposing a Special Topic. This set of elective courses are currently being reorganized into groupings that support specializations in the MCIS programme. The development of specializations provides coherence within the degree as it continues to grow, allowing new courses to be developed that align with the research interests of staff that complement existing courses.

Whilst the mixed background of the student cohort is a continuing challenge in terms of the provision of appropriate courses and learning strategies, it is also an excellent opportunity to adopt a peer learning approach that allows students to share their understanding with each other. For example, in Net-centric Computing students collaborate in project based activities [7]; in Information Security, students share ideas and provide feedback through a series of interactive presentations [8]. It is with this in mind that the SRE course has been designed, with a view to ultimately bridging the research-practice gap. The attempt to bring research and practice together in a single programme of study is consistent with the original aims and philosophy of the programme [9] and in line with the constructivist educational beliefs of the teaching team [10, 11]. In a broader context the issue addressed is the well acknowledged gap between rapidly evolving industry expectations and the traditionally slow responding academic curriculum.

## 3. Research-practice gap in RE

Stating user requirements has been a thorny issue since before the time of Brooks [12]. The difficulty of this was again stressed in 1990's by Hsia, Davis, & Kung [13], and it still remains a challenge today. Although techniques used in requirements engineering have been a research topic for some time, the results of such research has not always been adopted by practitioners. The state of practice is requirements are often still written in natural language, despite the drawbacks of such a representation. Processes and tools produced by researchers to aid practice may not be adopted by practitioners for various reasons. It has been observed that "without more technology transfer, RE practice is unlikely to improve and much RE research will remain irrelevant" [14].

Davis & Hickey [15] argue that many requirements engineering researchers fail to understand current practices or the actual needs of practitioners. They conclude, citing Redwine & Riddle [16], that "When we as requirements researchers lament that technology transfer takes a whopping 15 years, perhaps we should look no farther than ourselves". It is in response to work such as this that we started a systematic survey of the research practice gap and incorporated findings into the delivery of software requirements courses.

### 3.1. Barriers to adoption

Previous work [6] has conducted an extensive literature review of the research-practice gap, specifically related to requirements specification activities. In the literature there are several reasons cited for the research-practice gap [6]. Although researchers may be aware of the needs of the practitioners, there are other issues that have been brought to light in the last two decades that limit the effectiveness of research in requirements engineering being transferred into practice. There are two key barriers to bridging the research-practice gap that can be addressed through the design of software requirements engineering course.

Firstly, it has been shown that there is little or no collaboration between researchers and practitioners [17, 18]. Therefore the inputs to research projects do not always reflect the issues of requirements engineering in practice. Even if researchers come up with a workable solution, the scalability of such research outputs is not covered [18] due to non availability of industrial data, hence data with industrial strength should be employed in research [17]. There has been much investigation in to what makes collaborative projects succeed or fail, and we subscribe to the concept of Reflective Systems Development (RSD) [19]. In RSD it has been observed that "the practitioner does not function as a mere user of a research output, but reveals to the reflective researcher their ways of thinking, or world view, that is brought to current practice, and draws on reflective research as an aid to his own reflection-in-action practice. Moreover, the reflective researcher cannot maintain distance from, much less superiority to, the experience of practice … he must somehow gain an inside view of the experience of practice" [20]. It is our belief that the sharing of world-views required to ensure that such collaboration is successful is dependent on effective communication and the development of a shared world view, rather than the projection of one's own world view on to another.

Secondly, another key barrier to adoption is that requirements engineering is not taught to any depth in many universities. Students have only some vague knowledge through software engineering. Hence there is a lack of well trained requirements engineers [18]. Armarego [21] has reviewed much of the recent literature that supports the assertion that formal education for requirements engineering is a major challenge for the next decade. It is to address this second barrier that the SRE course was introduced. The design and delivery of the course is such that it address the first challenge, namely enabling students to gain a shared "world view" through peer learning activities and effective collaboration.

## 4. Software requirements engineering

### 4.1. Course overview

The Software Requirements Engineering (SRE) course has been run successfully for two iterations. The main aim of the course is to provide students with knowledge of the interdependencies between enterprise stakeholders, processes, and software systems and enable them to apply appropriate techniques to the gathering, analysis, documenting, and managing of software systems requirements throughout the complete lifecycle. The secondary aim is to promote a focus on organizational context and ensures that advanced technical goals and functions will be aligned with improved business performance through the development of high quality software systems. The explicit learning outcomes of the course are:

1. Explain the need to engineer and manage software requirements from a business perspective
2. Integrate requirements engineering activities into the software development process
3. Evaluate requirements elicitation techniques and justify the choice of techniques appropriate for the given organizational context
4. Reinforce organizational interdependencies throughout the process of client needs identifications
5. Generate models of requirements using a variety of notations and techniques in conjunction with different software development methodologies
6. Identify commonly used industry standards and incorporate these in the preparation of software requirement specifications
7. Critique and validate requirements by facilitating specification reviews

As with all MCIS elective courses, SRE is a 15 point course with an expected time commitment of 150 hours. Of these 150 hours, only 21 are allocated as formal contact time. Students are expected to manage the remaining time whilst taking part in self-paced reading, online discussions and completion of the required assessment items.

### 4.2. Teaching and learning strategies

In line with the overall philosophy of the MCIS programme [9] this course emphasizes project-based learning in which students are engaged and active learners. As a result their learning experience is one of personal transformation. The idea is to develop a constructive learning environment that values the practical application of knowledge and promotes critically reflective researchers and professionals with strong technical capabilities in the computing discipline. This is augmented by group exercises and a group assignment to enhance the opportunities for peer learning.

The learning environment and assessment programme are underpinned with a constructivist viewpoint where learning about RE practice and theory is actively constructed by students through social interactions and making sense of their environment. Students are encouraged to collaboratively test new ideas, research and theories against their existing mental models of practice by providing opportunities and new experiences to challenge their current knowledge. In this way the students incrementally bridge the research-practice gap as new cognitive structures, attitudes and concepts develop from their previous practical experiences and knowledge. This approach emphasizes the personal nature of knowledge construction, the social and collaborative aspects of knowledge sharing and the need for reflection on actions and decisions to internalize new knowledge [10].

**4.2.1. Lectures.** Given the diverse mix of students who typically enroll in the course, short lectures are useful to ensure that key principles are discussed and that a framework for shared understanding is in place. Lectures are typically short, no more than one hour, and followed by group activities (see section 4.2.2) that reinforce learning by allowing students to apply the principles to practical examples. The first lecture of the course provides a high level overview of the whole requirements engineering process. Subsequent lectures drill down in particular areas that correspond to stages with in the process, e.g. requirements elicitation, requirements specification etc.

**4.2.2. Group exercises.** Group exercises that follow on from a lecture are the key mechanism for encouraging peer learning. Students are challenged to apply what has been learnt in a dynamic way, often involving discussion or role play scenarios. Students are encouraged to critique each other's work when appropriate. Typically, group exercises relate to known circumstances or systems that allow all students to participate. A specific example is asking small groups to develop solution-independent requirements for an Automatic Teller Machine. At the start of the exercise a typical requirement written by a student would be "The user shall be able to enter a PIN number", however following discussion it soon becomes clear that a better requirement is "The user shall be able to verify their identity".

**4.2.3 Online discussion and activities.** A number of online activities are conducted throughout the duration of a semester using AUTonline, a Blackboard© based learning environment. These activities vary from semester to semester, but typically can include participation in discussion forums, challenge exercises, contribution to wikis, analyzing and critiquing requirements and so forth.

**4.2.4. Assignments.** The course incorporates two summative assignments that are the only form of assessment. The first assignment is conducted individually, whilst the second assignment is a group assignment. Since the inception of the course, the first assignment has been focused on analyzing the research issues in requirements engineering and the second has been a practice based task. In the most recent run of the course, during the first semester of 2008, the focus of both assignments was the research-practice gap in requirements engineering. By explicitly requiring students to investigate "research concerning practice" and then to demonstrate "practice informed by research", the implicit aim of the course to create a shared understanding has been achieved.

For the first assignment, students selected a sub-topic (e.g. requirements specification and description, requirements traceability, requirements prioritization etc) and undertook a review of the research literature in this sub-topic with a view to analyzing current research to identify critical differences with current practice. The assignment was designed to appeal to the more research focused students, who are used to undertaking literature reviews and undertaking critical analysis. Whilst the assignment was individual, students were encouraged to discuss their thoughts and findings in lecturer-led class room discussions.

The second assignment was a group based project, working for a simulated client (a member of academic staff) to develop a set of requirements for a perceived software need. This second assignment was designed to appeal to the more practice focused students, however groups were carefully formed to ensure that all groups had a mix of student types. The assessment criteria for the project included an element of demonstrating that the techniques used for generating the requirements set were not only effective, but also based in part on recent research literature. The difference in world views of research focused and practice focused students led to ongoing discussions of how to resolve the conflict between producing a complete set of internally consistent requirements, whilst also demonstrating practice being informed by research. A final element of this second assignment was the individual student presentations, where students were encouraged to share what had been learned regarding both practice and research throughout the course.

**4.3. Role of the lecturer**

A key aim of the delivery of the course is the construction of a student-centered learning environment. The role of lecturer moves away from that of a traditional teacher, to a much more multi-faceted role. The primary facet of the role is that of learning facilitator, though flexibility to adapt to include other identifiable facets is important to support the changing nature of the course delivery [22]. To achieve this flexibility, the lecturer must come to understand the meaning of students' ideas rather than just correct them [23]. Apart from the short lectures that introduce key material, the SRE course is very much focused on problem-based learning (PBL) which is an instructional technique in which students learn through solving problems and reflecting on their experiences [24].

One of the aspirations of the course is to generate an environment where both student and teacher construct the learning agenda. A key element of this construction is a continuous dialogue. Questioning is often used to guide student thinking. A particular technique (or style of questioning), gleaned from educational literature [25], is used - the reflective toss. The purpose of the reflective toss is to allow the lecturer to interpret the meaning of a student statement but ensure that the student continues to elaborate their underlying thinking.

The goal of the lecturer-as-facilitator is to move the focus of student learning away from simply remembering facts, towards some form of higher learning, such as the understanding of underlying principles. Such a goal is appropriate for Master's level students who should be able to demonstrate competency at the higher level skills of analysis, synthesis and evaluation.

## 5. Discussion

One of the key aims for the development of the SRE course has been to provide a mechanism to close the research-practice gap in requirements engineering. Whilst this goal will take considerable time to realize, there has been some realization of interim goals.

As an emerging University, the move away from just addressing the educational needs of Information Technology (IT) professionals to also providing an infrastructure for growing research capability has its own unique challenges. In essence, there is a second gap between the aspirations of students enrolled in programmes of study and the need of lecturers to grow research.

By designing assignments that investigate the "research into practice" and also "practice informed by research", it is possible to ensure a flexible delivery that is accessible to students who are either current industry professionals or those who have aspirations for further study and research. Combining such a mix of students in a single class has great benefit in terms of peer learning. Current practitioners can challenge research, whilst students wishing to pursue research in this area can gain a understanding of the needs of practitioners and build relationships that could potentially lead to successful collaborations.

To date, the SRE course has been successfully delivered in full, in 2008 and 2009. For each year, the paper was only offered in the first semester (typically March to July). The first run of the course only had 7 students enrolled, which makes it difficult to draw any significant conclusions from results or student feedback. Student feedback was collected using AUT's Student Evaluation of Papers (SEP) process, operated by the Institutional Research Unit. Such evaluations are an opportunity for students to give anonymous feedback on their perceptions of a specific paper; they are also an opportunity to communicate plaudits and concerns.

SEP surveys can have up to15 quantitative questions (8 fixed and 7 optional) and 4 qualitative questions (2 fixed and 2 optional). Students rate their level of satisfaction with each quantitative question on a seven-point scale and provide comments in relation to the qualitative questions. The fixed quantitative questions relate to areas that research has shown are critical and also relate to AUT policy on what should be evaluated. The 7 optional questions can be selected from a list of standard questions.

In the first run of the course, all students achieved well above a minimum passing grade and generally evaluated their experiences as positive. The second run of the course had an enrollment of 28 students. Of these, 2 did not submit all of the assignments and thus did not complete the course. The majority of the remaining students achieved a grade greater than a B-. Once again, the students completed an evaluation survey administered by a neutral member of administrative staff of the course. In total, 19 students took part. 72% of the respondents indicated that they would recommend the course to someone else. 74% indicated that the course generated both interest and challenge.

Their subjective responses in the survey included comments such as "*I have wanted to learn about requirements engineering and this course gave me more knowledge of the subject area*" and "*Good mix between theory and practical parts*". Students expressed appreciation of the strong support received throughout the course. The strengths of the course are in its coherent structure and the opportunities it provides for student centered learning. Personal engagement of staff, potentially publishable student work and strong links with ongoing research are the key success factors associated with effective computer science education, and it is hoped that these factors will become the hallmarks of the course.

## 6. Conclusions

This paper has provided a high level summary of a relatively new Master's level course in Software Requirements Engineering (SRE). The course was born out of the recognition that a research-practice gap exists in SRE. The course addresses the two key causes for the research-practice gap, namely a lack of effective communication and the relatively light coverage of requirements engineering material in University programmes. Explicitly tailoring assessments to focus on the research-practice gap has been found to be a useful mechanism to promote the long term removal of said gap. Students with a research-focus have been encouraged to share and communicate with students who have more of a practice focus. In addition, full time academic staff can reflect on the process of students gaining shared understanding as a means to improving their interaction with industrial research partners.

Directions for further work include the development of a complementary course tentatively named "Software Architecture" that will focus on the transition of requirements into a design framework and working software, as well as developing more rigour in the process of determining whether papers are meeting their prescribed goals. Such work may also help meet the predicted increased demand for highly qualified specialists in software engineering [26].